\input harvmac
\overfullrule=0pt
\def\Title#1#2{\rightline{#1}\ifx\answ\bigans\nopagenumbers\pageno0\vskip1in
\else\pageno1\vskip.8in\fi \centerline{\titlefont #2}\vskip .5in}

%
\Title{\vbox{\baselineskip12pt
\hbox{gr-qc/9801096}\hbox{HUTP-98/A006}}}
{\vbox{\centerline {Statistical Entropy of }
\centerline {De Sitter Space } }}


\centerline{Juan Maldacena and Andrew Strominger}

\bigskip\centerline{Department of Physics}
\centerline{Harvard University}\centerline{Cambridge, MA 02138}

\def\slc{SL(2,C)}

\vskip .3in

\centerline{\bf Abstract}
Quantum gravity in 2+1 dimensions with a positive cosmological
constant can be represented as an SL(2,C) Chern-Simons 
gauge theory. The symmetric vacuum of this theory is a degenerate configuration
for which the gauge fields and spacetime metric vanish, while de Sitter 
space corresponds to 
a highly excited thermal state. Carlip's approach to 
black hole entropy can be adapted in this context to 
determine the statistical entropy of 
de Sitter space. We find that it equals one-quarter the area 
of the de Sitter horizon, in agreement with the 
semiclassical formula. 
\smallskip
\noindent 

\Date{}

\lref\dj{S. Deser and  R. Jackiw,  
 Annals Phys. {\bf 153} (1984) 405.}
\lref\ba{M. Ba\~nados and A. Gomberoff, gr-qc/9611044.}
\lref\jbmh{J.\ D.\ Brown and M.\
Henneaux, {\sl Comm. Math. Phys.} {\bf 104} (1986) 207.}
\lref\at{A.\ 
Achucarro and P.\ Townsend, 
{\sl Phys. Lett.} {\bf B180} (1986).}
\lref\ew{E.\ Witten, {\sl
Nucl. Phys.} {\bf B311} (1989) 46.} 
\lref\clp{S.\ Carlip, Phys. Rev. {\bf D51} (1995) 632.}
\lref\ascv{A.\ Strominger and C.\ Vafa, {\sl Phys. Lett.} {\bf B379}
(1996) 99.}
\lref\dps{M.\ Douglas, J.\
Polchinski and A.\ Strominger, hep-th/9703031.}
\lref\jm{J.\ Maldacena,
hep-th/9711200.}
\lref\sh{S.\ Hyun, hep-th/9704005.}
\lref\bph{H.\ Boonstra, B.\ Peeters, and K.\ Skenderis, hep-th/9706192.}
\lref\ss{K.\ Sfetsos and K.\ Skenderis, hep-th/9711138.}
\lref\btz{M.\ Ba\~nados, C.\ Teitelboim and J.\ Zanelli, {\sl
Phys. Rev. Lett.} {\bf 69} (1992) 1849.}
\lref\bhtz{M.\ Ba\~nados, M.\ Henneaux, C.\ Teitelboim, and J.\ Zanelli, {\sl
Phys.Rev.} {\bf D48} (1993) 1506.}
\lref\chv{O.\ Coussaert, M.\ Henneaux and P.\ van Driel, Class. 
Quant. Grav {\bf 12} (1995) 2961.}
\lref\giha{G. W. Gibbons and S. W. Hawking, Phys. Rev. {\bf D15} (1977) 2738.}
\lref\ch{O.\ Coussaert and M.\ Henneaux, Phys. Rev. Lett {\bf 72} (1994) 183.}
\lref\chtwo{O.\ Coussaert and M.\ Henneaux, hep-th/9407181.}
\lref\nha{J. W. York, Phys. Rev. {\bf D28} (1983) 2929.}
\lref\rez{B. Reznik, Phys. Rev. {\bf D51} (1995) 1728.}
\lref\nhb{W. H. Zurek and K. S. Thorne, Phys. Rev. Lett. {\bf 54} (1985)
          2171.}
\lref\nhc{ J. A. Wheeler, {\it A Journey into Gravity and Spacetime}
Freeman, N.Y. (1990).}
\lref\nhd{ G. 't Hooft, Nucl. Phys. {\bf B335} (1990) 138.}
\lref\nhe{L. Susskind, L. Thorlacius and R. Uglum, 
Phys. Rev. {\bf D48} (1993) 3743.}
\lref\nhf{V. Frolov and I. Novikov, Phys. Rev. {\bf D48} (1993) 4545.}
\lref\nhg{M. Cvetic and A. Tseytlin, Phys. Rev. {\bf D53} (1996) 5619.}
\lref\nhh{F. Larsen and F. Wilczek, Phys. Lett. {\bf B375} (1996) 37.}
\lref\hw{G. T. Horowitz and D. Welch, Phys. Rev. Lett. {\bf 71} (1993) 328.}
\lref\gth{G. 't Hooft, gr-qc 9310026.}
\lref\lss{L. Susskind, J. Math, Phys. {\bf 36} (1995) 6377.}
\lref\ls{D. A . Lowe and A. Strominger, Phys. Rev. Lett {\bf 73} (1994) 1468.}
\lref\jc{J. A. Cardy, Nucl. Phys. {\bf B270} (1986) 186.}
\lref\scct{S. Carlip and C. Teitelboim, Phys. Rev. {\bf D51} (1995) 622. }
\lref\ms{G. Moore and N. Seiberg, Phys.Lett. {\bf B220} (1989) 422.}
\lref\eel{S. Elitzur, G. Moore, A. Schwimmer and 
N. Seiberg, Nucl.Phys. {\bf B326} (1989) 108.}
\lref\ewslc{E. Witten, Comm. Math. Phys. {\bf 137} (1991) 29.}
 
An inertial observer in the 2+1-dimensional 
de Sitter vacuum detects thermal radiation 
at the temperature \giha\
\eqn\tds{T={1\over 2 \pi \ell}.} 
The boundary of the past light cone of the observer's world line is 
the de Sitter horizon. The horizon area is $2\pi \over \sqrt{ \Lambda}$, 
where $\Lambda$ is the cosmological constant. The 
corresponding thermodynamic entropy is 
\eqn\dent{S={\pi \over 2 \sqrt{ \Lambda}G},}
where $G$ is Newton's constant. The microscopic origin 
of \tds\ and \dent\ is an outstanding mystery. 

In this paper we will shed some light on this mystery in 
the context of pure 2+1 gravity \dj . Our approach will be a direct 
adaptation to the de Sitter case of Carlip's analysis \clp\ of the 
2+1 black hole. 2+1 gravity with a positive cosmological constant 
is equivalent to an \slc\ Chern-Simons gauge theory \refs{\at,\ew}.
We consider this theory on a spatial disc whose boundary 
corresponds to the de Sitter horizon. After imposition of 
the horizon boundary conditions proposed in \clp, the 2+1 \slc\ gauge theory 
reduces to an \slc\ WZW theory on the horizon boundary of the 
disc \refs{\ew,\ms,\eel}. The ground state of this WZW theory 
has vanishing metric and gauge fields, and corresponds to the elusive 
state of unbroken symmetry in 2+1 gravity. De Sitter space arises as 
an excited
thermal state in which the metric acquires 
an expectation value. In this paper we show that the statistical 
entropy of this state agrees with the thermodynamic entropy \dent.

Carlip's derivation rests on several key assumptions. 
One is the existence of a suitably defined conformal field theory 
with the SL(2,R)$\times$SL(2,R) (which in our case of positive 
$\Lambda$ becomes SL(2,C)\foot{The SL(2,C) theory might be defined
as the
complexification of an SU(2) theory. The \slc\ Chern-Simons 
theory was extensively
studied as such in \ewslc, but the status of the theory on the disc
remains unclear.}) current algebra with the required properties. In particular 
the asymptotic growth of states should depend on the central charge 
in the familiar manner \jc\ dictated by unitarity, despite the fact that it is 
not a unitary theory.  A second concerns the
nature of the boundary conditions. The choice in \clp\ does not appear to be
unique, and other choices do not yield the desired value for the entropy. 
In this paper we shall not question any of 
these assumptions. We will find that 
in the positive $\Lambda$ context they do imply the de Sitter entropy \dent.
The validity of the assumptions certainly warrants further investigation.

2+1 gravity with a positive cosmological constant is 
described by the action  \eqn\ctn{I_{grav}={1 \over
 16\pi G} \int d^3 x \sqrt{-g} (R -{2\over \ell^2}) ,} where 
the cosmological constant is $\Lambda = {1\over \ell^2}$
and we have omitted surface terms. 
In the following we will be interested in a semiclassical limit,
which requires that the cosmological constant is small in Planck 
units, or equivalently
\eqn\lcs{\ell \gg G.}

\ctn\ has the de Sitter solution 
\eqn\adst{ds^2= -\left(1-{r^2\over \ell^2}  \right) dt^2 +
\left( 1-{r^2\over \ell^2 }\right)^{-1} dr^2 + r^2 d\phi^2,} 
where $\phi$
has period $2\pi$ and $0 \le r \le \ell$. Spacelike surfaces of 
constant time $t$ are discs $D_2$. The coordinate system 
\adst\ covers the 
portion of de Sitter space corresponding to the past light cone of 
an observer following the geodesic $r=0$. The boundary of 
the region of 
de Sitter space from which no causal signal can reach such 
an observer is the horizon at $r=\ell$. We seek a microscopic derivation 
of \dent\  in the context of the quantum theory of 
2+1 gravity on $D_2 \times R$.

The quantization of \ctn\ is best achieved by recasting 
the theory as an \slc\ Chern-Simons gauge theory \refs{\at,\ew}.
The action for this this theory is, up to boundary terms, 
\eqn\csac{I_{grav}[A]
={is \over 4 \pi}\int_{D_2 \times R} Tr\bigl( A\wedge dA 
+{2 \over 3} A \wedge A \wedge A \bigr) -{is \over 4 \pi}
\int_{D_2 \times R} Tr\bigl(\bar A\wedge d\bar A 
+{2 \over 3} \bar A \wedge \bar A \wedge \bar A\bigr) ,}
where, in the conventions of \clp\ ~\foot{The gauge group is a 
complexification of SL(2,R) with generators and 
structure constants normalized as 
$Tr T^a T^b=2\eta^{ab}$ 
 and the line element is $ds^2=2\eta_{ab}e^ae^b$.} 
\eqn\krd{s={\sqrt{2} \ell \over 8 G},}
and $A$ is an
\slc\ gauge field. \csac\ is equivalent to \ctn\ with the 
identification
\eqn\aew{A^a={\omega^a } +{i \over \ell}e^a,}
for $a=0,1,2$. Here $e^a=e^a_\mu dx^\mu$ is the triad 
and $\omega^a=\half \epsilon^{abc}\omega_{\mu bc}dx^\mu$ is the 
SL(2,R) spin connection. It is easily checked that the 
de Sitter solution \adst\ corresponds to 
\eqn\adso{\eqalign{A^0&=
\sqrt{{1\over 2}-{r^2\over 2 \ell^2}}(-d\phi+{i \over \ell}dt),\cr
                   A^1&= {i \over \sqrt{2\ell^2-{2r^2}}}dr, \cr
                   A^2&={r \over \sqrt{2} \ell^2}dt+{ir \over \sqrt{2}\ell} d\phi .\cr}}

The Chern-Simons theory \csac\ has no local dynamics. The quantum 
theory on $D_2 \times R$ is defined by imposing boundary conditions 
on the cylindrical $(\phi, t)$ boundary at
$r=\ell$ which set half of the 6 complex tangential 
components of the gauge field to 
fixed values $A^B$ \refs{\ew,\ms, \eel}. We wish to choose 
boundary conditions appropriate to a horizon of proper length $2\pi \ell$. 
Carlip \clp\ 
has argued that the appropriate boundary conditions are\foot{An 
alternate and simplified version of the 
boundary conditions can
be found in  \ba .}
\eqn\adsot{\eqalign{A^{B0}_\phi+A^{B1}_\phi&=0,\cr
                   A^{B0}_t+A^{B1}_t&= 0, \cr
                   A^{B2}_\phi &={\omega \over \sqrt{2}} 
+{i\over \sqrt{2}}.\cr}}
This insures that the induced metric on the horizon agrees with 
\adst\ and in particular is of proper length $2\pi \ell$. 
$\omega$ here is an arbitrary real constant. All constant values of 
$\omega$ correspond to a horizon on the boundary  
in some gauge \clp. In the following we will find the 
value of $\omega$ which gives the dominant contribution to the entropy. 

Imposing \adsot\ reduces \csac\ to a complex level $k=is$ \slc\ 
WZW conformal field theory 
on the boundary of the disc \refs{\ew,\ms, \eel,\clp, \ba}, after 
redefining $A\to \tilde A$ by a certain gauge transformation \clp .
The currents are defined by the expansion of  
$\tilde A_\phi$ on the boundary
\eqn\ajk{\tilde A_\phi={i \over s}\sum j_ne^{in\phi}.}
The \slc\  current algebra is
\eqn\calg{\eqalign{ [j_m^a, j_n^b]&=i{\epsilon^{ab}}_cj^c_{m+n}
+ism{1 \over
2}
\eta ^{ab}\delta_{m+n,0},\cr [j_m^a, \bar j_n^b]&=0,\cr
[\bar j_m^a, \bar j_n^b]&=i{\epsilon^{ab}}_c\bar j^c_{m+n}-
ism{1 \over 2}
\eta ^{ab}\delta_{m+n,0}.\cr}}
The Virasoro generator $L_0$ is then given by the Sugawara construction as
\eqn\lko{L_0={1 \over 2is-1}Tr\sum^{n=\infty}_{n=-\infty }: j_nj_{-n}:
-{1 \over 2is+1}Tr\sum^{n=\infty}_{n=-\infty }: \bar j_n\bar j_{-n}:.}
Writing the nonzero mode piece as level number $N$ this is 
\eqn\loh{\eqalign{L_0&=N+{1 \over 2is-1}Trj_0^2-{1 \over 2is+1}Tr\bar
j_0^2\cr
 &=N-{1 \over 1+4s^2}Tr(j_0^2+\bar j_0^2)- 
{2is\over 1+4s^2}Tr(j_0^2-\bar j_0^2).}}
The squares of the 
zero modes of $j$ are determined by the boundary conditions \adsot\
and their relation \ajk\ to $\tilde A$\foot{The $A \to \tilde A$ 
redefinition does not affect this relation.}:
\eqn\jsg{Trj_0^2=-s^2(\omega+i)^2.} 
Inserting this relation yields
\eqn\lkj{L_0=N+{2s^2(\omega-2s)^2\over 1+4s^2}-2s^2.}
The Wheeler - de Witt equation implies $L_0=0$ \clp. The maximal value of 
$N$ is then attained  for $\omega=2s$ as 
\eqn\vct{N=2s^2={\ell^2 \over 16 G^2}.}
In the semiclassical regime of large $s$ the central charge may be 
approximated by $c \sim 6$. The unitarity relation for the 
logarithm of the number of states satisfying \vct\ is to leading order in $s$
\eqn\uti{S=2 \pi \sqrt{c N \over 6}={\pi \ell \over 2 G},}
in agreement with the semiclassical formula \dent.

A computation of the de Sitter entropy in a fully dynamical theory such as 
string theory remains an outstanding challenge. 

\centerline{\bf Acknowledgements}
This work was supported in part by DOE grant DE-FG02-96ER40559.
We are grateful to S. Carlip for correspondence.

\listrefs

\bye